\documentclass[lettersize,journal]{IEEEtran}
\usepackage{amsmath,amsfonts}
\usepackage{algorithmic}
\usepackage{algorithm}
\usepackage{array}
\usepackage[caption=false,font=normalsize,labelfont=sf,textfont=sf]{subfig}
\usepackage{textcomp}
\usepackage{stfloats}
\usepackage{url}
\usepackage{verbatim}
\usepackage{graphicx}
\usepackage{cite}
\usepackage{booktabs}
\usepackage{hyperref}
\usepackage{colortbl}

\begin{document}

\title{Near-Orthogonal Overlay Communications in LoS Channel Enabled by Novel OAM Beams without Central Energy Voids: An Experimental Study}
\author{Yufei Zhao,~\IEEEmembership{Member, IEEE}, Xiaoyan Ma,~\IEEEmembership{Member, IEEE}, Yong Liang Guan,~\IEEEmembership{Senior Member, IEEE}, Yile Liu, Afkar Mohamed Ismail, Xiaobei Liu, Siew Yam Yeo, and Chau Yuen,~\IEEEmembership{Fellow, IEEE}
\thanks{This work was supported by the National Research Foundation, Singapore and Infocomm Media Development Authority under its Future Communications Research \& Development Programme, Grant No. FCP-NTU-RG-2022-011, and No. FCP-NTU-RG-2022-020. It was also supported by Temasek Laboratories@NTU seed research projects, No. TLSP23-13 amd No. TLSP24-05. (Corresponding author: Xiaoyan Ma)}
\thanks{Yufei Zhao, Xiaoyan Ma, Yong Liang Guan, Yile Liu, Afkar Mohamed Ismail, and Chau Yuen are with School of Electrical and Electronic Engineering, Nanyang Technological University, 639798, Singapore (e-mail: \{yufei.zhao, xiaoyan.ma, eylguan\}@ntu.edu.sg, yliu129@e.ntu.edu.sg, \{afkar.mi, chau.yuen\}@ntu.edu.sg).}

\thanks{Xiaobei Liu and Siew Yam Yeo are with the Temasek Lab@NTU, 639798, Singapore (e-mail: \{xpliu, ysiewyam\}@ntu.edu.sg).}

\thanks{This work has been submitted to IEEE Internet of Things Journal.}}


\maketitle

\begin{abstract}
This paper introduces a groundbreaking Line-of-Sight (LoS) Multiple-Input Multiple-Output (MIMO) communication architecture leveraging non-traditional Orbital Angular Momentum (OAM) beams. Challenging the conventional paradigm of hollow-emitting OAM beams, this study presents an innovative OAM transmitter design that produces directional OAM beams without central energy voids, aligning their radiation patterns with those of conventional planar wave horn antennas. Within the main lobe of these antennas, the phase variation characteristics inherent to OAM beams are ingeniously maintained, linking different OAM modes to the linear wavefront variation gradients, thereby reducing channel correlation in LoS scenarios and significantly augmenting the channel capacity of LoS-MIMO frameworks. Empirical validations conducted through a meticulously designed LoS-MIMO experimental platform reveal significant improvements in channel correlation coefficients, communication stability, and Bit Error Rate (BER) compared to systems utilizing traditional planar wave antennas. The experiment results underscore the potential of the novel OAM-based system to improve current LoS-MIMO communication protocols, and offer both academic and engineering guidance for the construction of practical communication infrastructures. Beyond its immediate contributions, this paper underscores a pivotal shift in the field of communications, pointing out that traditional communication algorithms have primarily focused on baseband signal processing while often overlooking the electromagnetic characteristics of the physical world. This research highlights that, in addition to radiation patterns, the wavefront phase variations of traditional antennas represent a new degree- of-freedom that can be exploited. Consequently, future communication algorithms designed around reconfigurable electromagnetic properties hold the promise of ushering wireless communication into a new era.

\end{abstract}


\section{Introduction}
In today's digital communication landscape, there is an urgent and escalating need for wireless transmission capabilities of higher capacity. This demand is driven by the explosive growth of data-intensive applications, including high-definition video streaming, virtual and augmented reality, and the burgeoning Internet of Things (IoT) \cite{IoT}. This surge places a significant strain on conventional communication infrastructures, which now must not only support higher data rates but also provide transmission methods that are both more reliable and efficient. This is particularly crucial in scenarios where Line-of-Sight (LoS) communication is dominant. LoS scenarios are commonplace in a variety of settings, ranging from urban cellular networks to satellite communications \cite{backhaul}. Here, the unobstructed path between transmitter and receiver presents a golden opportunity for achieving high-capacity transmissions. Nonetheless, such environments also pose distinct challenges, including pronounced channel correlation and constrained spatial diversity. These factors can significantly degrade the effectiveness of traditional Multiple-Input Multiple-Output (MIMO) systems \cite{correlated}.

The introduction of Orbital Angular Momentum (OAM) in Electro-Magnetic (EM) waves marks a significant leap forward in wireless communication \cite{NTT,twist}. Initially established within optical contexts \cite{USA,USA2}, the recognition of OAM's utility at radio frequencies opened avenues for vastly expanding EM communication channel capacities, spurring extensive research and development activities \cite{twist,TVT,yufei-EL,Chao1}. This transition from a theoretical concept in optical communications to a promising foundation in Radio Frequency (RF) applications marks a remarkable journey in the evolution of communication research \cite{Allen}\cite{Thide}. OAM's ability to significantly boost wireless communication systems through enhanced data throughput and advanced multiplexing techniques represents a noteworthy advancement \cite{Tamburini}. By enabling the transmission of multiple, orthogonally distinct data streams over the same frequency band, OAM significantly augments spectral efficiency with low calculation complexity \cite{NTT2}. Ongoing investigations continue to unveil OAM's potential across various applications, including satellite communications \cite{Chao1}, quantum information processing \cite{chao-magazine}, and radar detection \cite{Liukang}, highlighting its broad applicability in future wireless communication systems.

While the theoretical benefits of OAM are considerable, its real-world deployment has encountered significant hurdles, particularly concerning beam divergence. Traditional OAM beams are characterized by pronounced divergence angles, leading to widespread energy dispersion, especially over extended transmission distances \cite{Liang,yufei2}. Mitigating this divergence is crucial for the successful integration of OAM into practical applications. Initial approaches to generating OAM beams, such as employing spiral phase plates and holographic methods, have laid the groundwork for OAM research \cite{twist,yufei-IoT}. However, the unique doughnut-shaped intensity profile and spiral phase distribution inherent to OAM signals pose significant challenges in establishing a receiving antenna within a constrained area, especially as the transmission distance increases. This limitation has made it extremely difficult to use a single antenna or an antenna array to fully receive the spiral phase \cite{Wenchi1,Wenchi2}.

Addressing the challenge of beam divergence necessitates innovative approaches to OAM beam generation and propagation. Inspired by advancements in plane spiral OAM technology \cite{xiong50}, in this paper, we introduce a novel OAM generator tailored to mitigate the key challenges associated with beam divergence. This innovation, termed as the new OAM beams devoid of central energy void, marks a significant advancement from traditional OAM generation approaches \cite{Chen1}. Distinctively, our cutting-edge OAM generator is engineered to emit directional cone-shaped beams, paralleling the functionality of traditional planar wave antennas. Moreover, it boasts a unique feature: the induction of linear phase variations across different gradients within the main lobe. This feature leverages the intrinsic properties of OAM modes to enhance wavefront manipulation capabilities, surpassing those of traditional planar wave systems with the directional radiation pattern.

This invention is a revolutionary advancement in communication systems. It is well known that MIMO systems exploit rich scattering landscapes to boost channel capacity. However, in LoS contexts, where channel correlation significantly increases, the multiplexing potential of LoS-MIMO systems faces substantial limitations \cite{backhaul,correlated}. As wireless communication technology leaps forward from 5G to newer generations, LoS-MIMO transmission becomes a typical scenario in point-to-point interactions and millimeter-wave (mmWave) communications. OAM beams, with their unique electromagnetic properties, offer constructive solutions for addressing the issues of highly correlated channels and low rank in LoS-MIMO scenarios \cite{Wenchi2}. In these conditions, our innovative OAM generater emerges as a optional alternative to orthodox planar wave antennas for LoS-MIMO communications. It infuses additional physical channel data through diverse wavefront phase modifications, effectively diminishing the correlation among traditional LoS channels and significantly elevating the multiplexing capacity. By integrating these novel OAM beams, we aim to overcome the limitations of highly correlated channels of conventional LoS-MIMO system, thereby enhancing the development of communication infrastructures to achieve more high-efficient wireless communications in LoS scenarios.


To clarify, the key contributions of this paper are summarized as follows:
\begin{itemize}
\item Firstly, our work introduces a pioneering OAM generator design that uniquely eliminates the central energy void typically observed in OAM transmissions. This method does not rely on traditional array beamforming techniques, instead, it uses only one port to excite cone-shaped radiation beams with OAM wavefront characteristics. By adjusting the waveguide dimensions, we successfully designed and fabricated various single-mode OAM generators. Notably, these generators were applied in MIMO communication experiments for the first time, marking a significant leap in LoS-MIMO applications.
\item Next, from the perspective of channel correlation, we modeled and simulated a typical LoS-MIMO system and compared changes in channel capacity when using traditional planar wave directional antennas and the novel OAM antennas proposed in this paper. The simulation setup closely replicates real communication environments, offering valuable insights for constructing actual physical experimental platforms.
\item Then, from theory to practice, we developed a comprehensive 2$\times$2 LoS communications experimental setup based on programable units and stable RF links. This experimental platform is equipped with a pre-defined signal frame structure that facilitates swift measurement of the multi-pair communications channel matrix. Additionally, it enables the assessment of channel correlation, thereby laying the foundational infrastructure for our comparative experiments.
\item Finally, leveraging the experimental platform, we conducted meticulous comparative measurements of the complex channel matrices under varying transmitter conditions¡ªspecifically, by alternating the Tx antennas between standard horn antennas and our novel OAM antennas. Our experiments unveiled that the application of our specially designed OAM antennas significantly reduces channel correlation and augments the channel capacity of LoS-MIMO systems. These empirical findings crucially validate the theoretical benefits of our innovative OAM antenna design in practical scenarios, underscoring the reduced channel correlation's impact on enhancing efficiency and reliability in high-demand communication environments, particularly in LoS-MIMO settings.
\end{itemize}

The rest of this paper is organized as follows. Section II covers the operational principles of the new OAM transmitter, along with simulation and measurement results. Section III explores how channel correlations affect channel capacity within the context of LoS-MIMO communication channels. Section IV demonstrates performance improvements in real communication experiments, supported by the actual experimental platform. Finally, Section V concludes this paper and discusses the implications of the findings, potential applications, and avenues for future research.

\section{Novel OAM Generator without Central Energy Void} \label{sec2}
\subsection{Theory, Design and Analysis}
Traditionally, the generation mechanism of OAM waves is predicated on the spatial phase distribution characteristics of EM waves. When the phase of an EM wave exhibits a continuous rotational change along a propagation direction within a closed loop, it results in a wave carrying OAM. Unlike uniform phase distributions, the phase of OAM wave varies in a helical manner with respect to the azimuthal position, completing a phase change of $2\pi \ell$ for each revolution ($\ell$ being an integer known as the topological charge or OAM mode number), indicating the wave carries $\ell$ units of orbital angular momentum. In the field of RF, the commonly used methods for generating OAM waves include the following well-known techniques, for example, 1) Spiral Phase Plates: By varying the thickness of the medium at different positions, EM waves passing through acquire a helical phase distribution, thereby carrying OAM; 2) Phased Array Antennas: By electronically controlling the phase of each element within an array antenna, EM waves with a helical phase distribution are generated. This is one of the most commonly used methods in the microwave and radio frequency domains. 3) Ring Antennas: Utilizing the physical characteristics of ring antenna structures, EM waves with specific OAM modes are generated through adjustments in feeding methods or antenna structural design.

In summary, the traditional mechanism for generating OAM waves involves a systematic process starting with phase rotation, where each segment of the EM wave demonstrates a continuous phase change along its propagation path, culminating in a uniform helical phase distribution. This process is distinguished by a topological charge, denoted by $\ell$, which represents the degree of phase shift per revolution around the propagation axis as an integer multiple of $2\pi$. This charge not only highlights the wave's rotational dynamics but also quantifies the orbital angular momentum inherent to the wave. The characteristic structure of these waves is observed in their beam profile, which typically features a ``doughnut'' shape in its transverse cross-section, marked by a central void representing zero intensity. This unique formation is indicative of the non-zero angular momentum that OAM waves possess \cite{theory1}. However, OAM waves generated based on this classical mechanism all share a common issue of a ``central energy void'' phenomenon within the beam, leading to energy divergence. This central energy void not only affects the operational range of traditional OAM beams, but also poses more stringent challenges to the receiver during multiplexing transmission.

\begin{figure}[htbp]
\centering
\includegraphics[width=3.4in]{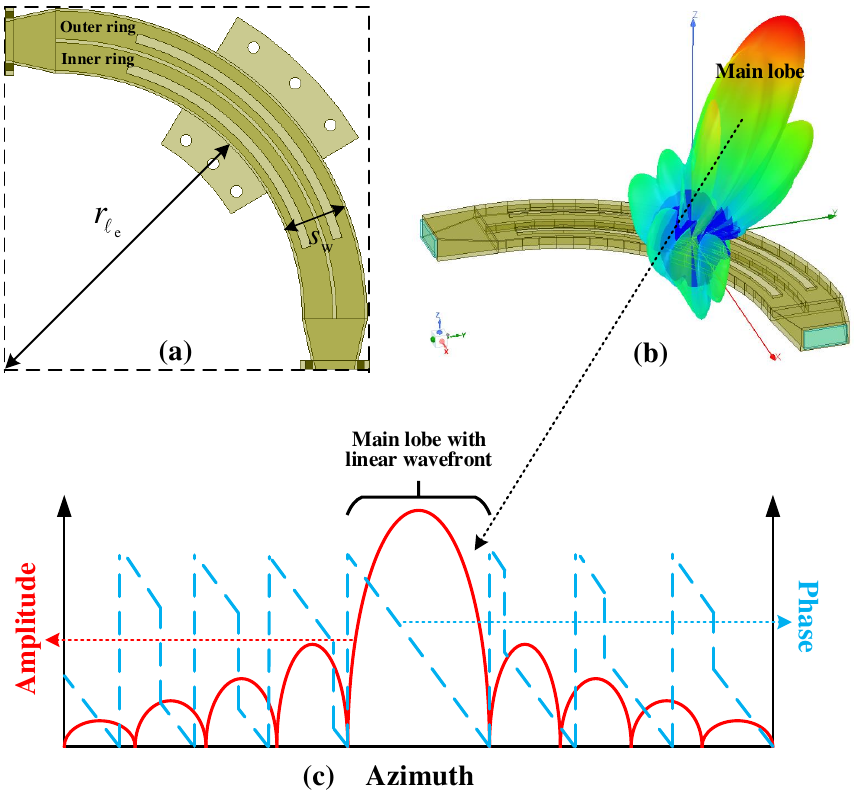}
\caption{A NTCS-OAM generator design without central energy void with respect to (a) Physical structure design (b) 3D radiation pattern with simulation model (c) 2D beam schematic diagram showing the amplitude (solid line) and the phase (dashed line) distributions. }
\label{wavefront}
\end{figure}

To eliminate the central energy void in traditional OAM beams, the principle of Fourier series expansion can be applied. This principle allows periodic complex exponentials to serve as bases for constructing various signals. In the spatial domain, as we know, the OAM waves exhibit an azimuthal phase distribution of $\exp \left( { - {\rm{j}}\ell\varphi } \right)$, and different OAM modes introduce periodic spatial phase variations. By superimposing these modes, we can achieve a far-field radiation pattern with arbitrary angular intensity and phase distributions \cite{xianmin}.
Recently, a method employing ``Non-Uniform Traveling-Wave Current Sources (NTCS)'' to generate OAM beams has been introduced \cite{zhu}. This approach, evolving from OAM mode-group theory \cite{Shilie}, refines the uniform traveling-wave source model to derive a method for distributing current sources in a non-uniform manner around a ring. This technique can produce EM waves with specific OAM characteristics.
Inspired by the previous research, this paper proposes a method for exciting OAM waves with high-gain and arbitrary modes. It is worth noting that the generated OAM beam's directional pattern exhibits high directivity, similar to that of traditional cone beams. This approach avoids the typical ``doughnut'' shape of traditional OAM beams, resulting in more concentrated energy and a propagation distance comparable to that of a plane directional beam with equivalent gain. Unlike plane waves, the novel OAM beams produced in our study exhibit a distinct linear phase variation within the main lobe. The gradient of this variation is determined by the topological mode number of the OAM beam, as shown in Fig. \ref{wavefront}.

In this paper, arc-shaped waveguides are utilized to simulate the NTCS generators. As depicted in Fig. \ref{wavefront}, one end of the waveguide is excited by a signal source while the other end is terminated with a matching load, thereby establishing a traveling-wave transmission environment within the waveguide. The waveguide comprises a wide side ${s_{\rm{w}}}$, and a narrow side ${s_{\rm{n}}}$, operating under the ${\rm{T}}{{\rm{E}}_{10}}$ fundamental mode. Consequently, by creating appropriate slot along the waveguide's wide side, a slotted leaky-wave radiation antenna can be constructed. According to the theory of traveling-wave waveguides, the relationship between the operating cutoff wavelength $\lambda _{\rm{c}}^{{\rm{T}}{{\rm{E}}_{10}}}$ of the waveguide and the cavity dimensions can be expressed as follows,
\begin{equation}
\lambda _{\rm{c}}^{{\rm{T}}{{\rm{E}}_{10}}} = \frac{{{\lambda _0}}}{{\sqrt {1 - {{\left( {{{{\lambda _0}} \mathord{\left/
 {\vphantom {{{\lambda _0}} {2{s_{\rm{w}}}}}} \right.
 \kern-\nulldelimiterspace} {2{s_{\rm{w}}}}}} \right)}^2}} }},
\end{equation}
where ${{\lambda _0}}$ denotes the wavelength in vacuum. Then, define the current source model $I_n(\phi) = I_{n0}(\phi)e^{-j\ell_n(\phi)\phi}$, where $I_{n0}(\phi)$ denotes the amplitude distribution of the current, and $\ell_n(\phi)$ represents a mode distribution function that varies with the arc angle $\phi$. According to \cite{zhu}, if the radiation source is a uniform traveling-wave current source, it will generate a single mode OAM beam which can be expressed as,
\begin{equation} \label{ntcs}
\vec U\left( {d,\theta ,\varphi } \right) = \alpha \frac{{{\mu _0}{I_0}f{e^{ - jkd}}}}{{2d}}{J_{{\ell _0}}}\left( {k{r_{\rm{c}}}\sin \theta } \right){e^{ - j{\ell _0}\varphi }},
\end{equation}
where, $\vec U\left( {d,\theta ,\varphi } \right)$ is the radiation electric field, $k$ is wave vector, $d$ is transmission distance, $f$ is frequency, ${{\mu _0}}$ is the permeability in free space, $\alpha $ is a constant propagation parameter, $\theta $ and $\phi $ denotes the pitch and azimuth independent variables of the radiation pattern, ${{r_{\rm{c}}}}$ denotes the radius of the radiation source, and ${J_{{\ell _0}}}\left(  *  \right)$ denotes the ${{\ell _0}}$-th Bessel function of the first kind. The construction of a NTCS-OAM generator based on \eqref{ntcs} is as follows. Based on the principle of Fourier series expansion, a NTCS can be regarded as the result of numerous uniform traveling-wave current sources after weighting. In other words, the radiation field of an NTCS in free space can also be equivalently considered as the superposition of multiple different mode OAM beams, that is,
\begin{equation} \label{eq1}
U\left( {d,\theta ,\varphi } \right) = {C_{\rm{t}}}\sum\limits_\ell  \begin{array}{l}
{\rm{Sinc}}\left[ {\frac{{{\varphi _{\rm{c}}}\left( {\ell  - {\ell _{\rm{e}}}} \right)}}{2}} \right]\\
 \times {J_\ell }\left( {k{r_{{\ell _{\rm{e}}}}}\sin \theta } \right){e^{ - j{\ell _{\rm{e}}}\frac{{{\varphi _{\rm{c}}}}}{2}}}{e^{ - j\ell \left( {\varphi  - {\varphi _{\rm{d}}}} \right)}}
\end{array},
\end{equation}
where, ${C_{\rm{t}}} = \alpha \frac{{{\mu _0}{I_0}2\pi f{e^{ - jkd}}}}{{4\pi d}}$, ${{\varphi _{\rm{c}}}}$ denotes the angle corresponding to the NTCS, ${{\varphi _{\rm{d}}}}$ is the radiation direction of the main lobe, ${\ell _{\rm{e}}}$ denotes the center OAM topological mode number within the mode spectrum. In practical applications, ${\ell _{\rm{e}}}$ can be obtained by measuring the linear slope within the main lobe of such novel OAM beams. Actually, there are various configurations of waveguide leaky-wave antennas \cite{zhu2}. In this paper, we employ the simplest ${90^ \circ }$ arc-shaped structure to generate directionally radiated OAM beams. To produce novel OAM beams of different topological modes ${\ell _{\rm{e}}}$, we can adjust the radius ${r_{{\ell _{\rm{e}}}}}$ of the arc-shaped waveguide. The relationship between them can be derived as,
\begin{equation}
{r_{{\ell _{\rm{e}}}}} = \frac{{\left| {{\ell _{\rm{e}}}} \right|}}{{\pi \sqrt {{{\left( {2/{\lambda _0}} \right)}^2} - {{\left( {1/{s_{\rm{w}}}} \right)}^2}} }} - \frac{{{s_{\rm{w}}}}}{2}.
\end{equation}

\subsection{Simulation, Fabrication and Measurement}
\begin{figure}[htbp]
\centering
\includegraphics[width=3.4in]{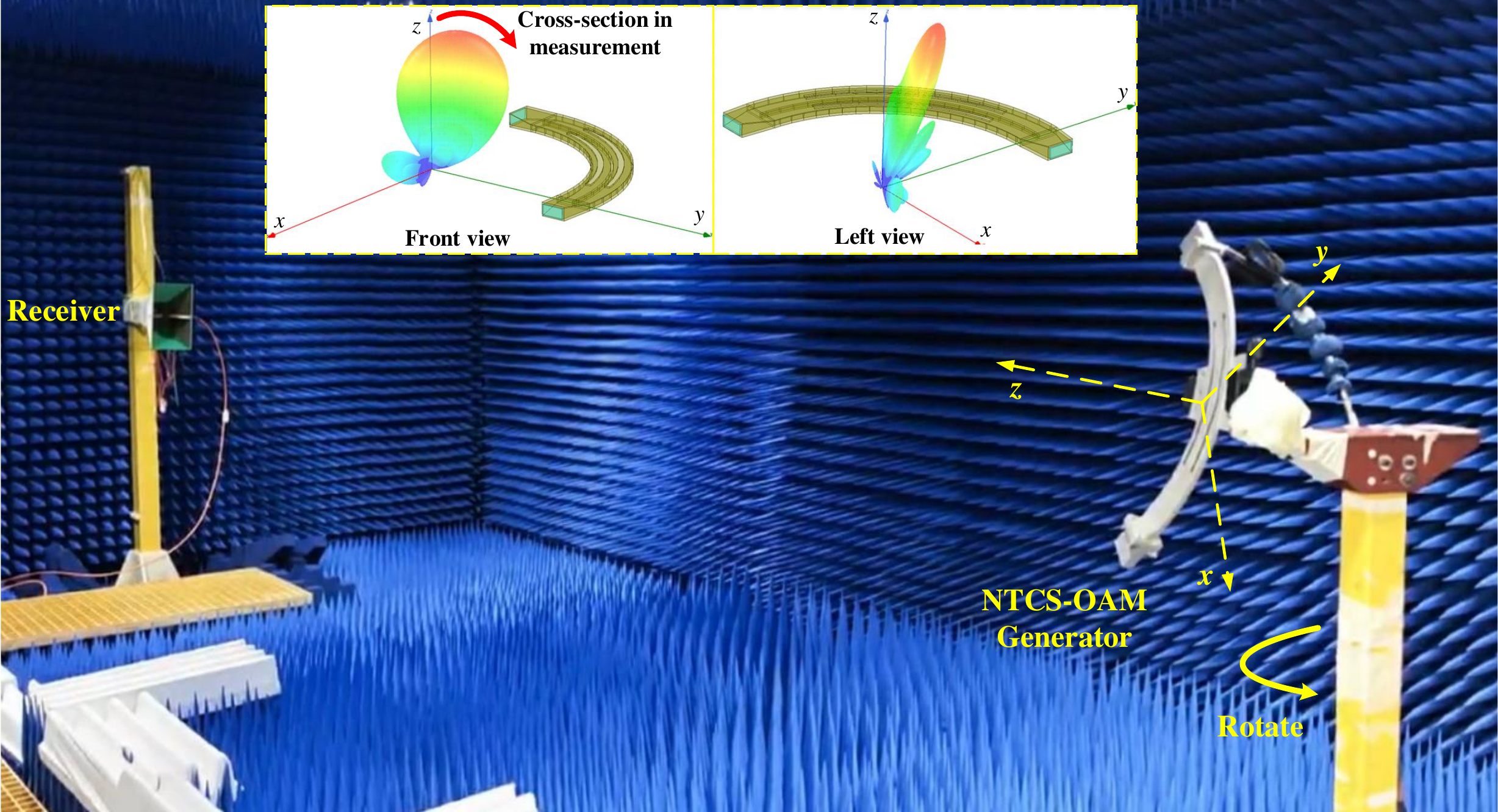}
\caption{Measurement scenario in the microwave anechoic chamber, refer video at: \href{https://youtu.be/e-PU-Td8_yc/}{YouTube Video}.}
\label{measurement}
\end{figure}

\begin{figure}[htbp]
\centering
\includegraphics[width=2.9in]{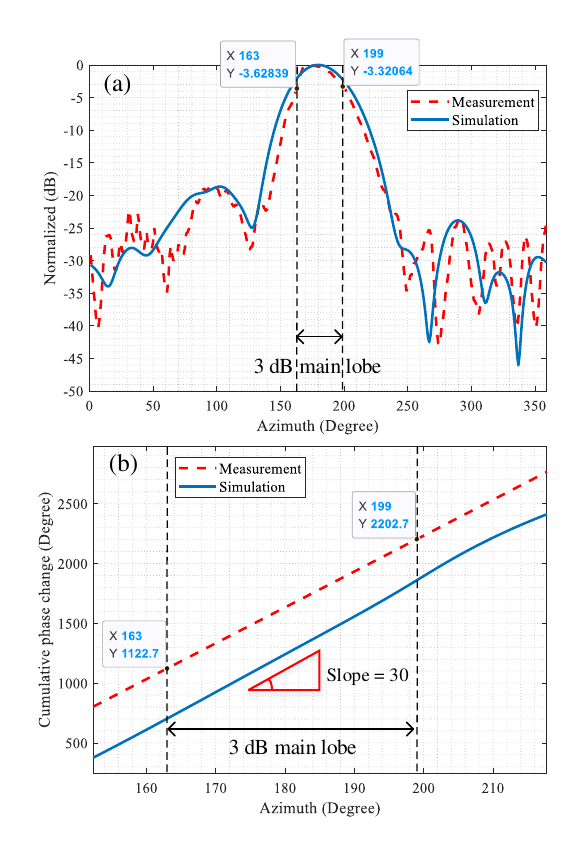}
\caption{Novel 2D radiation pattern of OAM mode ${\ell _{\rm{e}}} = 30$, comparing results between simulation (solid line) and measurement (dashed line).}
\label{OAM1}
\end{figure}

\begin{figure}[htbp]
\centering
\includegraphics[width=2.9in]{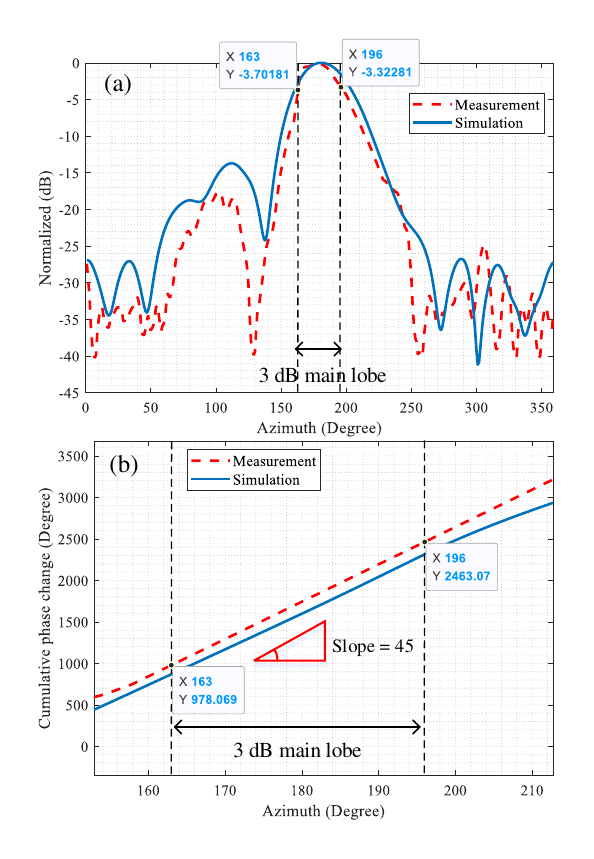}
\caption{Novel 2D radiation pattern of OAM mode ${\ell _{\rm{e}}} = 45$, comparing results between simulation (solid line) and measurement (dashed line).}
\label{OAM2}
\end{figure}

To enhance the radiation gain of the leaky-wave antenna, parallel slots were cut on the wide side of the waveguide, as shown in Fig. \ref{wavefront}(a)(b). They share the same waveguide feed port and load matching port. Significantly, the radiation pattern generated by this NTCS-OAM antenna exhibits a conical shape, similar to that of traditional planar wave antennas. However, a notable distinction lies within the main lobe of the radiation pattern, as shown in Fig. \ref{wavefront}(c), where the variations in the wavefront phase of the beam are designed independently. The simulation process was carried out in Ansys HFSS software, and the results are illustrated in Fig. \ref{OAM1} and Fig. \ref{OAM2}. The measurement scenario is shown in Fig. \ref{measurement}. In the microwave anechoic chamber, we sequentially mounted each OAM transmitter of varying modes onto a turntable, working at 10 GHz. Following simulation guidelines, we tilted the partial arc NTCS-OAM antenna by approximately ${18^ \circ }$ and allowed it to rotate ${360^ \circ }$ around its geometric center.

Traditionally, researchers in communications field often regards antennas primarily as directional energy radiation devices. In practical applications, the focus has typically been on inherent metrics such as radiation direction, antenna gain, and operating frequency bandwidth. However, it is crucial to recognize that EM waves, as a vectorial form of physical manifestation, possess not only specific amplitude characteristics but also variations in wavefront phase within the beam \cite{sha-wei}. OAM represents a form of structured EM wave characterized by a helical wavefront phase. Through specialized antenna design, we have successfully eliminated the central energy void typically associated with traditional OAM beams. This approach concentrates the radiated energy in a single direction, forming a conical beam while retaining the unique wavefront phase characteristics of OAM, manifested as linear phase variations within the main lobe of the radiation pattern. As shown in Fig. \ref{OAM1} and Fig. \ref{OAM2}, both the amplitude envelopes and phase distributions within the 3 dB main lobe of the measurement results closely match the simulation curves. Clearly, within the 3dB beamwidth of the main lobe, the wavefront phase varies linearly, and the slope of the measured curve aligns well with the simulation results. The slope of the wavefront phase change is determined by the equivalent OAM topological mode number. For instance, from the measurements in Fig. \ref{OAM1}, we can calculate that the equivalent OAM topological mode number of the directed OAM beam radiated by the NTCS antenna is ${\ell _{\rm{e}}} = \left( {{{2202}^ \circ } - {{1122}^ \circ }} \right)/\left( {{{199}^ \circ } - {{163}^ \circ }} \right) = 30$. Similarly, from Fig. \ref{OAM2}, we derive that the equivalent mode number is ${\ell _{\rm{e}}} = \left( {{{2463}^ \circ } - {{978}^ \circ }} \right)/\left( {{{196}^ \circ } - {{163}^ \circ }} \right) = 45$, indicating that the OAM beam exhibits a steeper phase change within the main lobe of a similar shape.

At this point, one might wonder why there is a need to construct such complex wavefront phase variations within directive beams that initially have nearly identical radiation patterns. Recent research \cite{Geng-Bo} suggests that this approach actually introduces a new dimension to beam shaping. Imagine, if we could programmatically control the wavefront phase structure within a traditional directional beam without altering its radiation pattern. This capability would create a novel method of EM control, independent of traditional beam steering techniques. This method, distinct from traditional beam steering techniques, could profoundly transform signal processing approaches in fields such as communication, detection, and localization. Next, we will analyze how the variation in wavefront phase can be utilized to reduce the correlation of physical channels under conditions of strong correlation assumption, thereby increasing the channel capacity of the entire communication system.

\section{Wireless Channel Capacity from the Correlation Perspective} \label{correllation}
It is well known that in wireless communication, there is a specific type of channel model known as the LoS transmission channel. LoS channel is extensively used in high-frequency wireless transmissions (such as terahertz and millimeter-wave frequency band) \cite{frequency}, satellite communications \cite{satellite}, and point-to-point backhaul scenarios \cite{backhaul}. Due to the lack of multipath scattering conditions in these communication environments, the propagation of EM wave signals through space is primarily affected by large-scale fading. This results in a situation where, when transmitting (or receiving) antennas are situated close to each other, the channel fading experienced by signals traveling from different transmitting antennas to the receiving end exhibits a high degree of correlation \cite{backhaul}.

For a long time, MIMO communication in LoS scenarios has presented a persistent and challenging issue for researchers. This is because the implementation of MIMO's multiplexing and diversity capabilities relies on a rich scattering environment \cite{correlated,Wenchi2}. However, in LoS scenarios, the lack of scattering conditions makes the parallel channels extremely similar, leading to significant correlation in the propagation paths of multiplexed signals. Under these circumstances, traditional pre-coding or channel equalization algorithms become ineffective. In other words, the rank of LoS-MIMO channels severely degrades, which in turn significantly impacts the overall channel capacity of the communication system. There was a time when OAM was considered a subset of MIMO. This was primarily because traditional methods of generating vortex beams relied on spatial beam shaping using array antennas, which led people to associate OAM with Uniform Circular Arrays (UCA) \cite{NTT,Chen2}. The debate over whether OAM represents a new dimension in wireless communication persisted for a long time. Currently, a widely accepted consensus is that under the assumption of generating OAM beams based on UCA, OAM may not offer a new dimension of beamforming beyond what MIMO provides. However, it can reduce the complexity of the transceiver system and is an effective method to achieve robust communication links.

In this paper, we introduce an OAM beam generation method based on NTCS, which is fundamentally different from the traditional UCA approach. If we consider a conventional MIMO system where each array element is a planar wave directional antenna, such as a horn antenna, and the entire MIMO system is composed of multiple such arrays, by the same token, we can replace the planar wave arrays in this MIMO system with the novel OAM antennas designed in this study. Consequently, each array element would no longer radiate a directional planar wave, but instead a directional OAM beam with a specially designed wavefront within the main lobe. Base on this assumption, we can present an in-depth analysis of the channel matrix in LoS-MIMO systems, focusing on the channel correlation coefficient and covariance matrix, and their impact on channel capacity. Let us consider a $m \times n$ channel matrix $\mathbf{H}$ with $m$ Tx antennas and $n$ Rx antennas in a LoS-MIMO system, represented as,
\begin{equation}
{\mathbf{H}} = \left[ {\begin{array}{*{20}{c}}
  {{h_{1,1}}}& \cdots &{{h_{1,m}}} \\
   \vdots & \ddots & \vdots  \\
  {{h_{n,1}}}& \cdots &{{h_{n,m}}}
\end{array}} \right] \in {\mathbb{C}^{n \times m}}.
\end{equation}
This matrix is fundamental in characterizing the transmission properties of the system. Assuming at the LoS-MIMO scenarios, the distance between the $m$-th Tx antenna and the $n$-th Rx antenna is denoted as ${d_{n,m}}$. Then, for the plane wave, the element in the channel matrix $\mathbf{H}$ can be derived based on the well know free space transmission theory as,
\begin{equation}
{h_{n,m}} = \beta \frac{\lambda }{{4\pi d}}{e^{ - jk{d_{n,m}}}},
\end{equation}
where $\beta $ is the channel attenuation coefficient related to the environment. Furthermore, according to \eqref{eq1}, if the Tx antennas are replaced by the special designed NTCS-OAM radiator, the channel element between each transceiver can be re-written as,
\begin{equation}
h_{n,m}^{{\rm{OAM}}}\left( {\theta ,\varphi } \right) = \beta \frac{\lambda }{{4\pi d}}{g_m}\left( {\theta ,\varphi } \right){e^{ - jk{d_{n,m}}}}{e^{ - j\ell _{\rm{e}}^m\varphi }},
\end{equation}
where, ${g_m}\left( {\theta ,\varphi } \right)$ denotes the radiation gain of the novel OAM beams at the $m$-th Tx antenna, and ${\ell _{\rm{e}}^m}$ is the equivalent topological mode. It should be noted that, to better align with the actual transmission scenario of the communication channels, both ${g_m}\left( {\theta ,\varphi } \right)$ and ${\ell _{\rm{e}}^m}$ can be measured by the full wave simulation through HFSS or CST, rather than directly derived from ideal theoretical model in \eqref{eq1}. Without loss of generality, taking a channel matrix with $2 \times 2$ transceivers as an example.
The receive correlation coefficient $\rho$ is a measure of the linear relationship between the received signals at different antennas. It can be calculated using the elements of $\mathbf{H}$ as,
\begin{equation}
\rho = \frac{ \left| h_{11} h_{21}^{*} + h_{12} h_{22}^{*} \right| }{ \sqrt{ ( |h_{11}|^2 + |h_{12}|^2 ) \times ( |h_{21}|^2 + |h_{22}|^2 ) } } \in \left[ 0,1 \right].
\label{correlation}
\end{equation}
The value of $\rho$ ranges between 0 and 1, indicating the degree of correlation. The receiving covariance matrix $\mathbf{R}$ represents the expected value of the matrix formed by multiplying $\mathbf{H}$ with its Hermitian transpose, which can be expressed as,
\begin{equation}
\begin{gathered}
  {\mathbf{R}} = \mathbb{E}\left[ {{\mathbf{H}}{{\mathbf{H}}^H}} \right] \hfill \\
  {\kern 1pt} {\kern 1pt} {\kern 1pt} {\kern 1pt} {\kern 1pt} {\kern 1pt} {\kern 1pt} {\kern 1pt} {\kern 1pt} {\kern 1pt} = \mathbb{E}\left[ {\begin{array}{*{20}{c}}
  {{h_{11}}h_{11}^* + {h_{12}}h_{12}^*}&{{h_{11}}h_{21}^* + {h_{12}}h_{22}^*} \\
  {{h_{21}}h_{11}^* + {h_{22}}h_{12}^*}&{{h_{21}}h_{21}^* + {h_{22}}h_{22}^*}
\end{array}} \right] \in {\mathbb{C}^{2 \times 2}}. \hfill \\
\end{gathered}
\end{equation}

Then, the channel capacity $C$, which measures the maximum data transmission rate with low error probability, is related to these parameters. It can be approximated as \cite{correlated},
\begin{align}
\begin{array}{*{20}{l}}
  C\!\!\!\!\!\!&{ \approx \log \det \left( {\mathbf{R}} \right) + \Delta } \\
  {}&{ \approx \log \left( \begin{gathered}
  ({h_{11}}h_{11}^* + {h_{12}}h_{12}^*) \times ({h_{21}}h_{21}^* + {h_{22}}h_{22}^*) \hfill \\
   - ({h_{11}}h_{21}^* + {h_{12}}h_{22}^*) \times ({h_{21}}h_{11}^* + {h_{22}}h_{12}^*) \hfill \\
\end{gathered}  \right) \!+\! \Delta } \\
  {}&{ = \log \left( {N(1 - {\rho ^2})} \right)\! +\! \Delta ,}
\end{array}
\end{align}
where $N = (|h_{11}|^2 + |h_{12}|^2) \times (|h_{21}|^2 + |h_{22}|^2)$ simplifies the notation, and $\Delta$ represents a portion of $C$ independent of $\mathbf{R}$. In the range of $\rho \in \left[ 0,1 \right]$, the channel capacity $C$ monotonically decreases with respect to $\rho$. This indicates that a reduced value of the receiving correlation coefficient $\rho$ is conducive to an enhanced achievable channel capacity $C$. This analysis demonstrates that the channel capacity in LoS-MIMO systems is influenced by the channel correlation coefficient and the covariance matrix, highlighting the importance of minimizing correlation in LoS-MIMO system design.

\begin{figure}[htbp]
\centering
\includegraphics[width=2.9in]{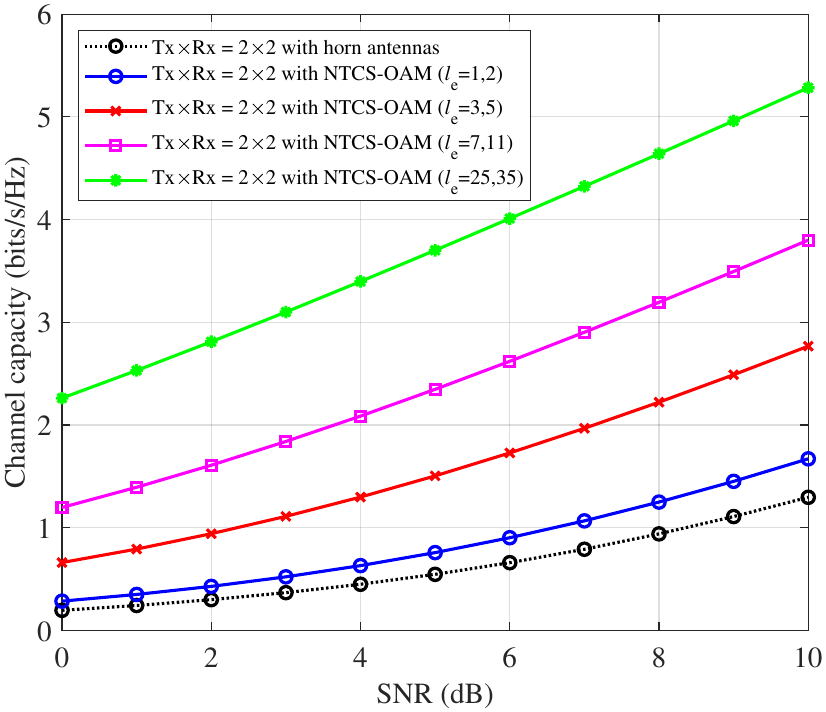}
\caption{Channel capacity of a 2-pair LoS communication scenario with different OAM modes (Frequency at 10 GHz, Tx/Rx antennas spacing is 0.2 m, LoS distance between Tx and Rx is 10 m, antenna height is 1.5 m, radiation power is 0 dBm).}
\label{capacity1}
\end{figure}

The channel capacity of a 2-pair LoS communication scenario with horn antenna and different modes NTCS-OAM antennas is shown in Fig. \ref{capacity1}. The simulation results in Fig. \ref{capacity1} indicate that due to the lack of scattering conditions in LoS channels, the proportion of multipath components between the transmitter and receiver is extremely low, leading to a high correlation within the channel. This is a practical issue commonly faced by traditional backhaul links. Taking a simulation setup with $2 \times 2$ transceivers as an example, when horn antennas are used at the Tx end, the EM field conditions of the two antennas are very similar, leading to severe degradation in the channel's rank. This degradation is evident in the channel capacity curve, where, despite increasing Signal-to-Noise Ratios (SNR), there is little substantial improvement in capacity. In contrast, when we replace the Tx antennas with the novel NTCS-OAM antennas proposed in this paper, there is a significant improvement in channel capacity. Moreover, as the difference in OAM modes between the two Tx increases, the channel capacity further enhances. This demonstrates that by adjusting the wavefront phase distribution within the main radiation lobe, it is possible to reduce the correlation among LoS channels, thereby improving the conditions for multiplexing communication in LoS-MIMO scenarios.

\begin{figure}[htbp]
\centering
\includegraphics[width=2.9in]{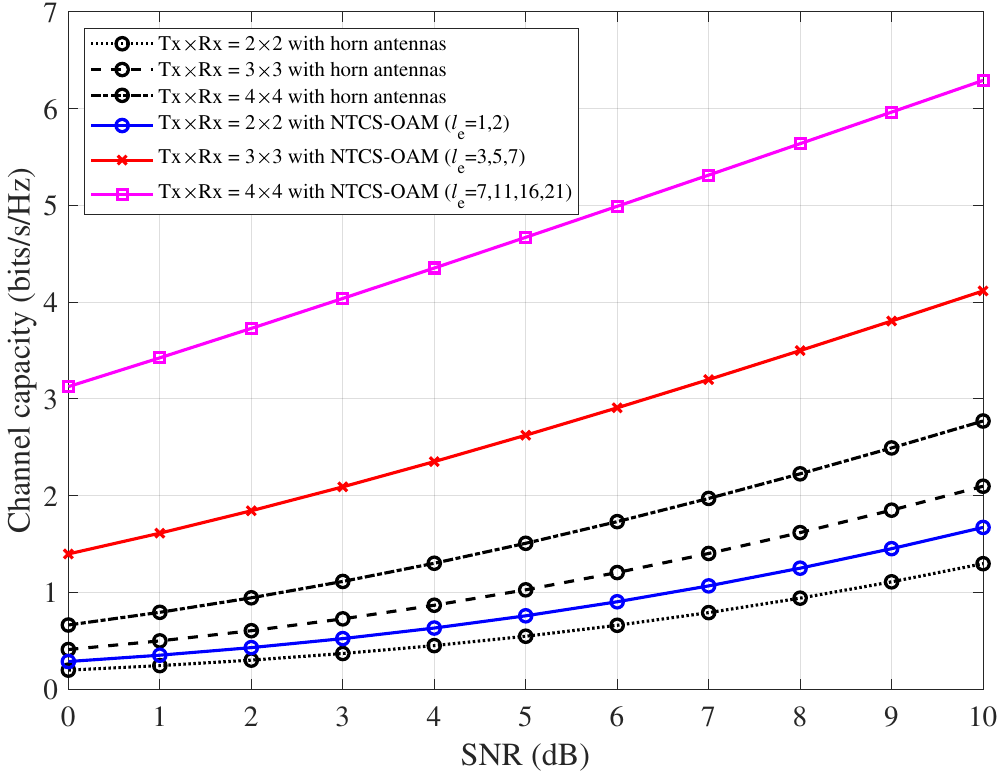}
\caption{Multi-pair LoS communication scenarios with or without OAM (Frequency at 10 GHz, Tx/Rx antennas spacing is 0.2 m, LoS distance between Tx and Rx is 10 m, antenna height is 1.5 m, radiation power is 0 dBm).}
\label{capacity2}
\end{figure}

Then, we sought to enhance system capacity by increasing the number of antennas. It is well-known that in traditional MIMO systems, the number of transceiver antennas is closely linked to the number of multiplexing channels, i.e., the rank of the wireless channel. As the number of antennas increases, the channel capacity should theoretically increase proportionally. However, due to the high correlation typical in LoS environments, using traditional antennas often fails to achieve a proportional increase in total system capacity. By switching to NTCS-OAM antennas, which allow for the manipulation of wavefront phase to reduce antenna correlation, we observed that channel capacity significantly improved even in LoS environments, which has been demonstrated in Fig. \ref{capacity2}. This improvement scales with the increase in the number of NTCS-OAM antennas, demonstrating that effective wavefront control can mitigate the typical correlation issues associated with LoS conditions and thus enhance system performance.

Furthermore, let's quantify this phenomenon from the perspective of the channel's rank and condition number. It is known that in MIMO systems, by performing Singular Value Decomposition (SVD), the singular values of the channel can be obtained, which represent the total number of channels that can be multiplexed for communication. The relative size of these singular values indicates the channel quality. Specifically, the condition number of the channel is defined as the ratio of the largest singular value to the smallest singular value, i.e.,
\begin{equation}
{\eta _{{\text{cond}}}} = \frac{{\max \nu _i^2}}{{\min \nu _i^2}},
\end{equation}
where $i \in \left\{ {1,2, \ldots ,{\rm{rank}}\left[ {\bf{H}} \right]} \right\}$, ${\max \nu _i^2}$ and ${\min \nu _i^2}$ denote the maximum and the minimum singular value of the channel matrix respectively. ${\eta _{{\text{cond}}}}$ can serve as a metric to evaluate the quality of multiplexing capabilities in a MIMO system. A lower condition number indicates a channel with more uniform singular values, which is desirable as it suggests that the channel has better and more balanced communication capacities across its multiplexed channels. Conversely, a high condition number indicates significant disparities in singular values, suggesting poor channel quality where certain channels dominate the communication capability, potentially leading to inefficiencies and imbalance in the use of the channel's capacities \cite{yufei-IoT}.

\begin{table}[htbp]
\caption{Channel conditional number ${\eta _{{\rm{cond}}}}$ for Fig. \ref{capacity1}.}
\begin{center}
\begin{tabular}{c|c|c|c|c}
\toprule
\text{Horns} & \textbf{${\it{l}}_{e}=1,2$} & \textbf{${\it{l}}_{e}=3,5$} & \textbf{${\it{l}}_{e}=7,11$} & \textbf{${\it{l}}_{e}=25,35$} \\
 \midrule
 $2.01 \times {10^{16}}$ & $106.11$ & $51.65$ & $25.62$ & $10.05$ \\
 \bottomrule
\end{tabular}
\end{center}
\label{tab_cond}
\end{table}
\begin{table}[htbp]
\caption{Channel conditional number ${\eta _{{\rm{cond}}}}$ for Fig. \ref{capacity2}.}
\begin{center}
\begin{tabular}{c|c|c|c}
\toprule
\textbf{Setups} & \textbf{Tx$\times$Rx: 2$\times$2} & \textbf{Tx$\times$Rx: 3$\times$3} & \textbf{Tx$\times$Rx: 4$\times$4} \\
 \midrule
 Tx with horns & $2.01 \times {10^{16}}$ & $4.29 \times {10^{32}}$ & $3.94 \times {10^{33}}$ \\
 Proposed NTCS & $1.06 \times {10^{2}}$ & $6.03 \times {10^{3}}$ & $2.55 \times {10^{17}}$ \\
 \bottomrule
\end{tabular}
\end{center}
\label{tab_cond}
\end{table}

Based on this, we have compiled the channel condition numbers from simulation Fig. \ref{capacity2} into Table \ref{tab_cond}. It is evident that under the same physical conditions, the channel condition numbers using traditional horn antennas are extremely high, indicating that increasing the number of antennas does not effectively result in multiplexing gains. On the other hand, the use of NTCS-OAM antennas significantly improves the condition numbers, leading to an increase in channel capacity that stems from this improvement. This demonstrates the effectiveness of NTCS-OAM antennas in reducing channel correlation and enhancing the multiplexing capabilities of the system, even in challenging LoS environments.

\section{Prototyping and Experiments}
\subsection{Setups for the Testbed}
\begin{figure}[htbp]
\centering
\includegraphics[width=3.5in]{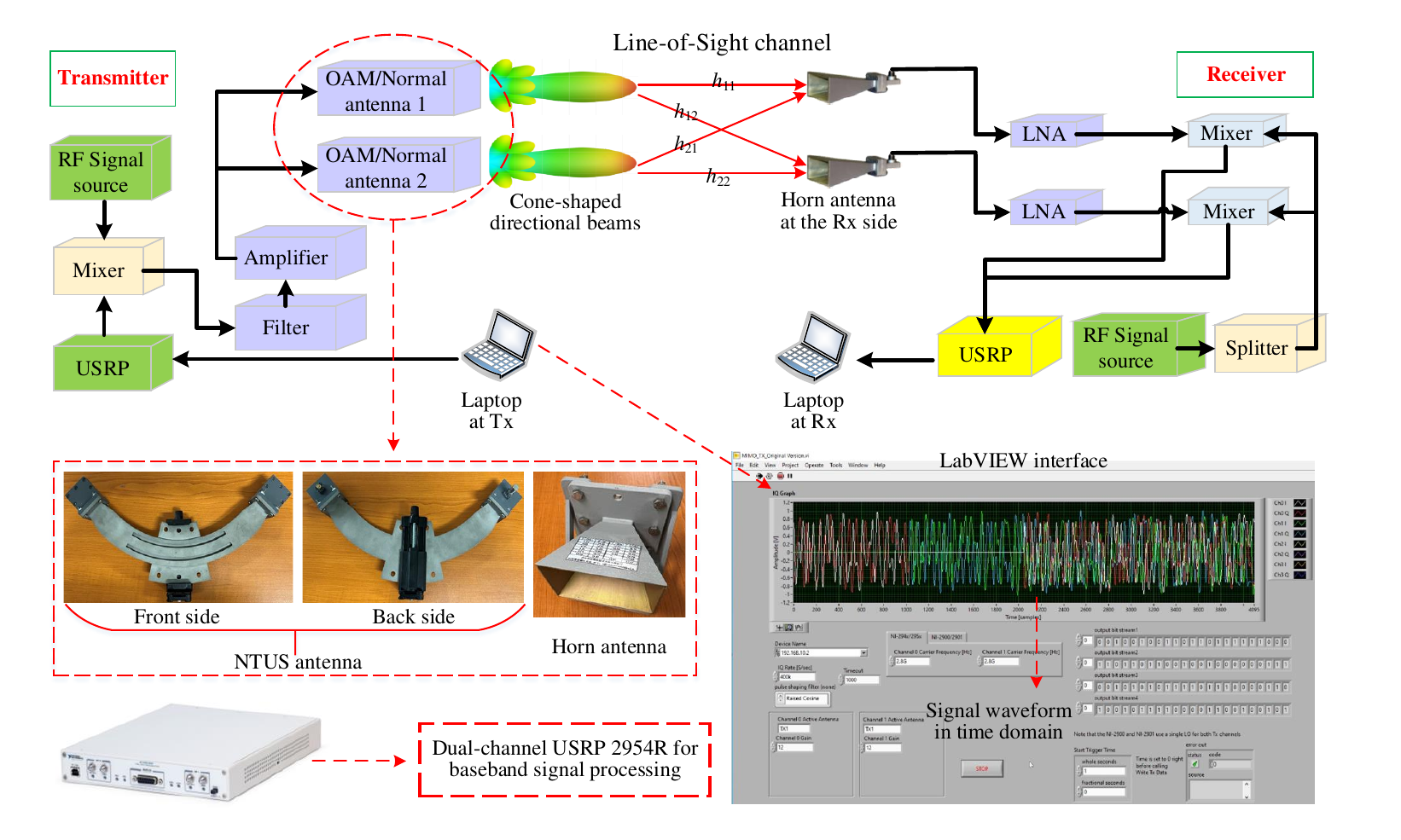}
\caption{Schematic diagram of the experiment system.}
\label{system}
\end{figure}

Based on the theoretical analysis presented in Sect. \ref{sec2}, we designed and fabricated two NTCS-OAM antennas, each emitting OAM beams of different modes without central void. The EM characteristics of these antennas, both simulated and measured, are displayed in Fig. \ref{OAM1} and Fig. \ref{OAM2}. At the receiving end, we utilized two identical horn antennas to establish a 2-transmitter, 2-receiver communication experiment platform. Specifically, two dual-channel Universal Software Radio Peripheral (USRP, NI USRP-2954R model) units were positioned at the Tx and Rx sides, serving as the core modules for baseband signal processing. The dual-channel modulated signals are up-converted and amplified before being fed into the NTCS-OAM antenna ports, and radiated into free space. At the Rx side, these RF signals are captured and down-converted to Intermediate Frequency (IF), after which the receiving USRP samples and demodulates them into two distinct baseband data streams. The modulated signals are mapped onto the received constellation diagram, and the demodulated bits are used for error rate statistics. The system diagram of the entire experimental platform is shown in Fig. \ref{system}. (The record video of the experiments can refer to: \href{https://youtu.be/cQiwJqzv7m4}{YouTube Video}.)

The central frequency of the RF signals radiated by the antennas is 10 GHz. The transmitter and receiver are approximately 2.5 meters apart, with no obstructions in the transmission path, thus creating a typical LoS communication environment. The main experiment parameters are listed in Table \ref{parameter}. The experimental platform uses LabVIEW software from NI company as its operating interface. Integrating USRP with LabVIEW simplifies the creation of advanced wireless communication systems for users. With LabVIEW, users can effortlessly manage the USRP device, handling signal transmission and reception, data analysis, and the development of intricate wireless algorithms, all while bypassing the complexities of low-level hardware intricacies.

\begin{table}[!h]
\caption{Main experiment parameters.}
\begin{center}
\begin{tabular}{c|c|c}
\toprule
\textbf{Parameter} & \textbf{Value} & \textbf{Dimension} \\
 \midrule
 Central carrier frequency & 10 & GHz \\
 Intermediate frequency & 2.8 & GHz \\
 I/Q Sampling Rate & 400 & kHz \\
 Samples per symbol & 16 & - \\
 Signal bandwidth & 25 & kHz \\
 Antenna gain & 16 & dB \\
 Tx antenna spacing & 1.2 & m \\
 Height of Tx/Rx & 1.5 & m \\
 Modulation for pilot signal & 16-QAM & - \\
 Modulation for data stream & 16-QAM & - \\
 \bottomrule
\end{tabular}
\end{center}
\label{parameter}
\end{table}

\subsection{Comparative Experiments Results}
\begin{figure}[htbp]
\centering
\includegraphics[width=3.4in]{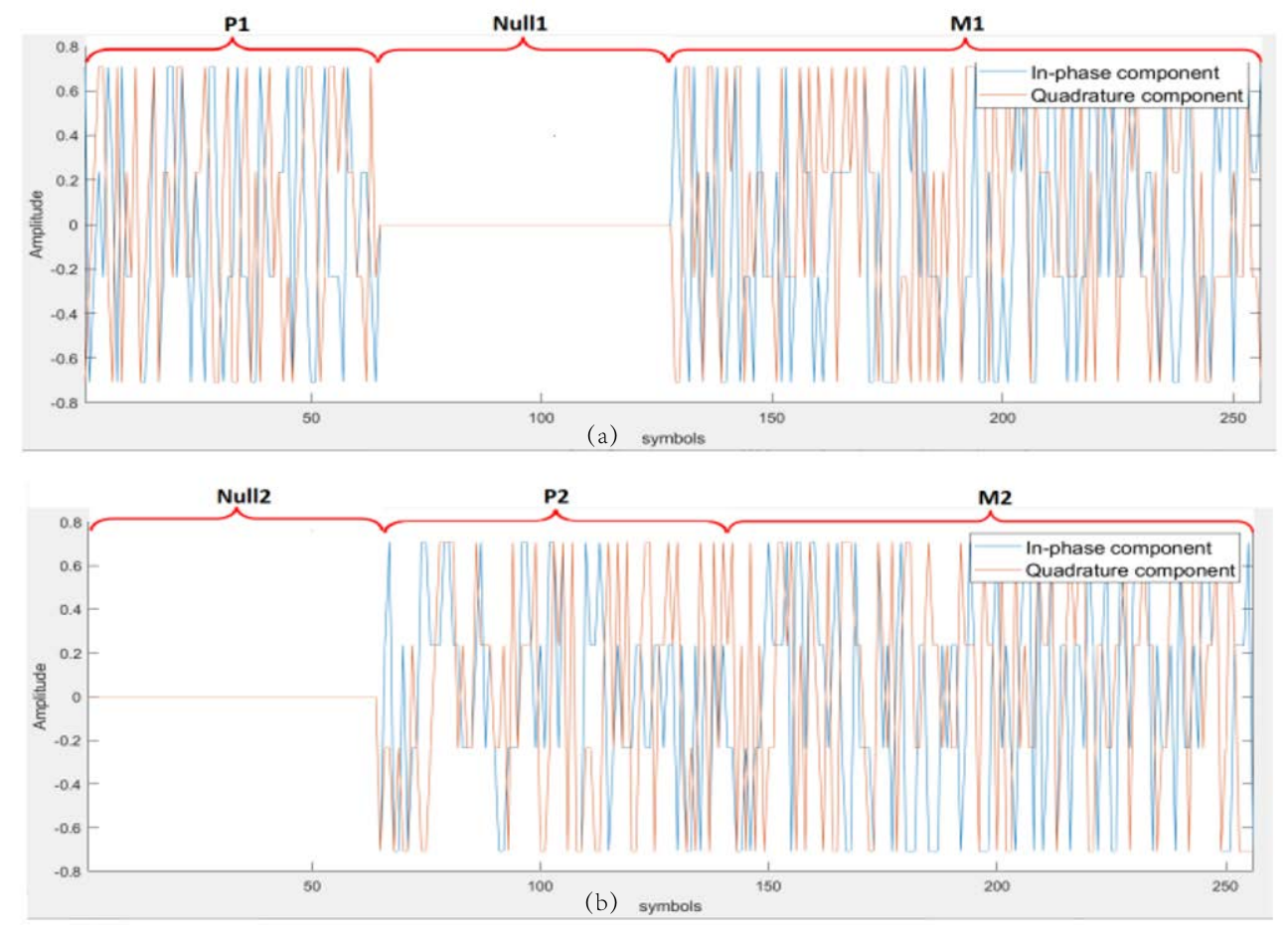}
\caption{Frame structure of the transmission signals with respect to (a) Data stream 1 and (b) Data stream 2.}
\label{frame}
\end{figure}

Fig. \ref{frame} shows the frame structure of the two channel multiplexing signals at the Tx side. Regardless of whether the transmitter uses NTCS-OAM antennas or regular plane wave horn antennas, the frame structure of their baseband signals is the same. During the experiments, 2 different data streams are transmitted at the same time. The channel from Tx antenna 1 to Rx 1 is $h_{11}$, which is used to send data stream 1, while the channel from Tx antenna 2 to Rx 2 is $h_{22}$, and it is used to send data stream 2. At the same time, Tx 1 and 2 will cause interference to each other, since they are sending signals at the same time using the same frequency, with the channels denoted by $h_{12}$ and $h_{21}$. The binary bits are modulated by 16-QAM, up-converted and fed to the antenna ports. $\textbf{P1}$ and $\textbf{P2}$ represent the training pilot signals for Tx 1 and 2, which are transmitted at the different time slots to guarantee the channel estimation quality. After receiving the pilot signals, the Zero Forcing (ZF) channel estimation will be performed at the receiver side through USRP, so that we can get the Channel State Information (CSI) to decode the message. $\textbf{M1}$ and $\textbf{M2}$ denote the binary messages that need to be transmitted by Tx 1 and 2 multiplexed at the same time slot.

\begin{figure}[htbp]
\centering
\includegraphics[width=3.4in]{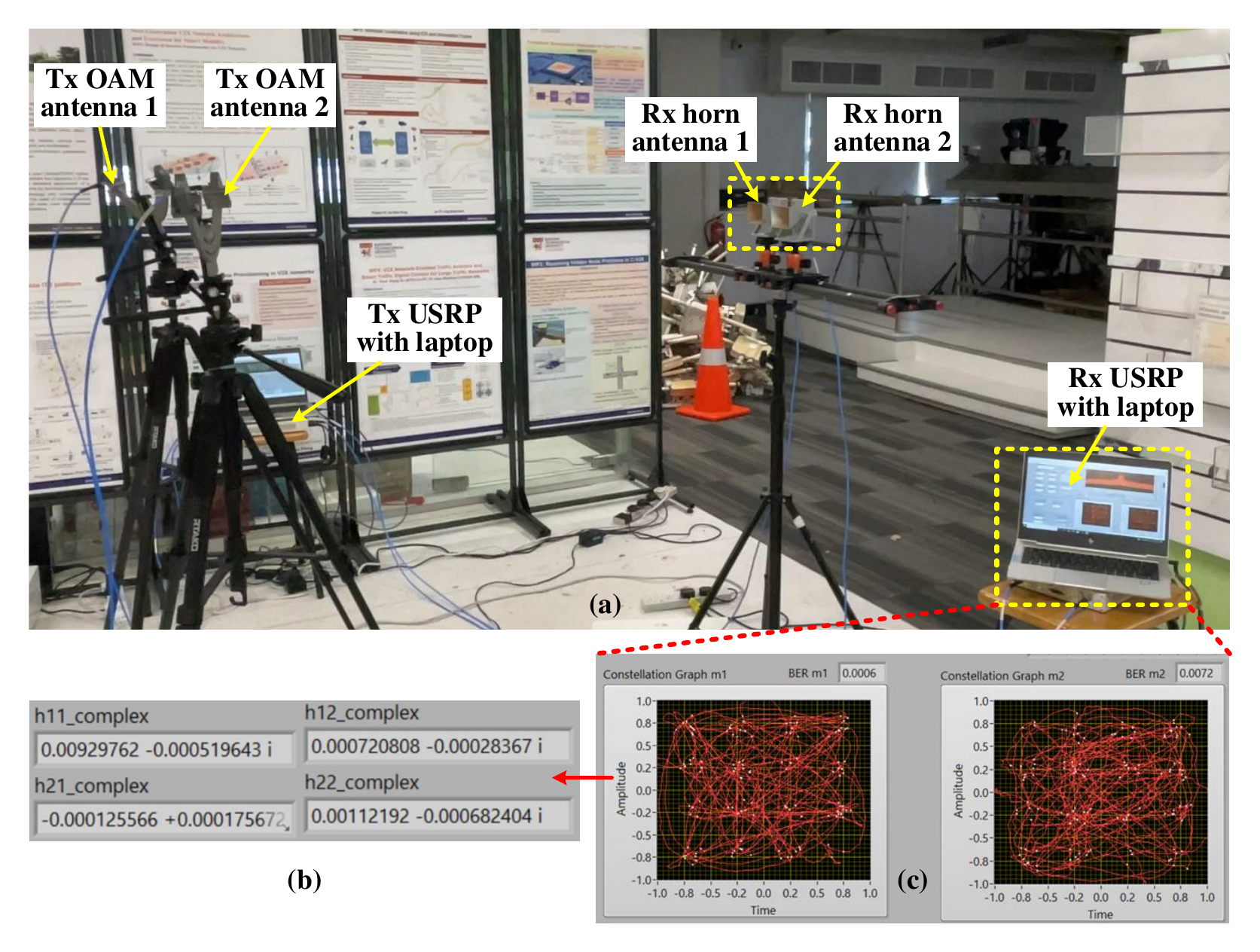}
\caption{Communications in high-correlation scenario with NTCS-OAM antennas at Tx side. (a) Experiment setups. (b) Measured channel matrix. (c) Demodulated constellations.}
\label{OAM-setup}
\end{figure}

The experiment setup is shown in Fig. \ref{OAM-setup}. Firstly, as shown in Fig. \ref{OAM-setup}, a $2 \times 2$ LoS-MIMO communication system has been established, featuring 2 NTCS-OAM antennas as the transmitters and 2 conventional horn antennas as the receivers. It is worth noting that the two receivers are positioned very close to each other, leading to high-correlation transmission scenarios. The receivers are connected to USRP so that we can see the real-time demodulated constellations. Without loss of generality, we recorded the instantaneous CSI, as shown in Fig. \ref{OAM-setup}(b), and used it to calculate the channel correlation coefficient $\rho _{{\rm{OAM}}}$ under the corresponding settings,
\begin{equation} \label{OAM_correlation}
{\rho _{{\rm{OAM}}}} = \frac{{\left| {{h_{11}}h_{21}^* + {h_{12}}h_{22}^*} \right|}}{{\sqrt {(|{h_{11}}{|^2} + |{h_{12}}{|^2}) \times (|{h_{21}}{|^2} + |{h_{22}}{|^2})} }}{\rm{ = }}0.1140.
\end{equation}

\begin{figure}[htbp]
\centering
\includegraphics[width=3.4in]{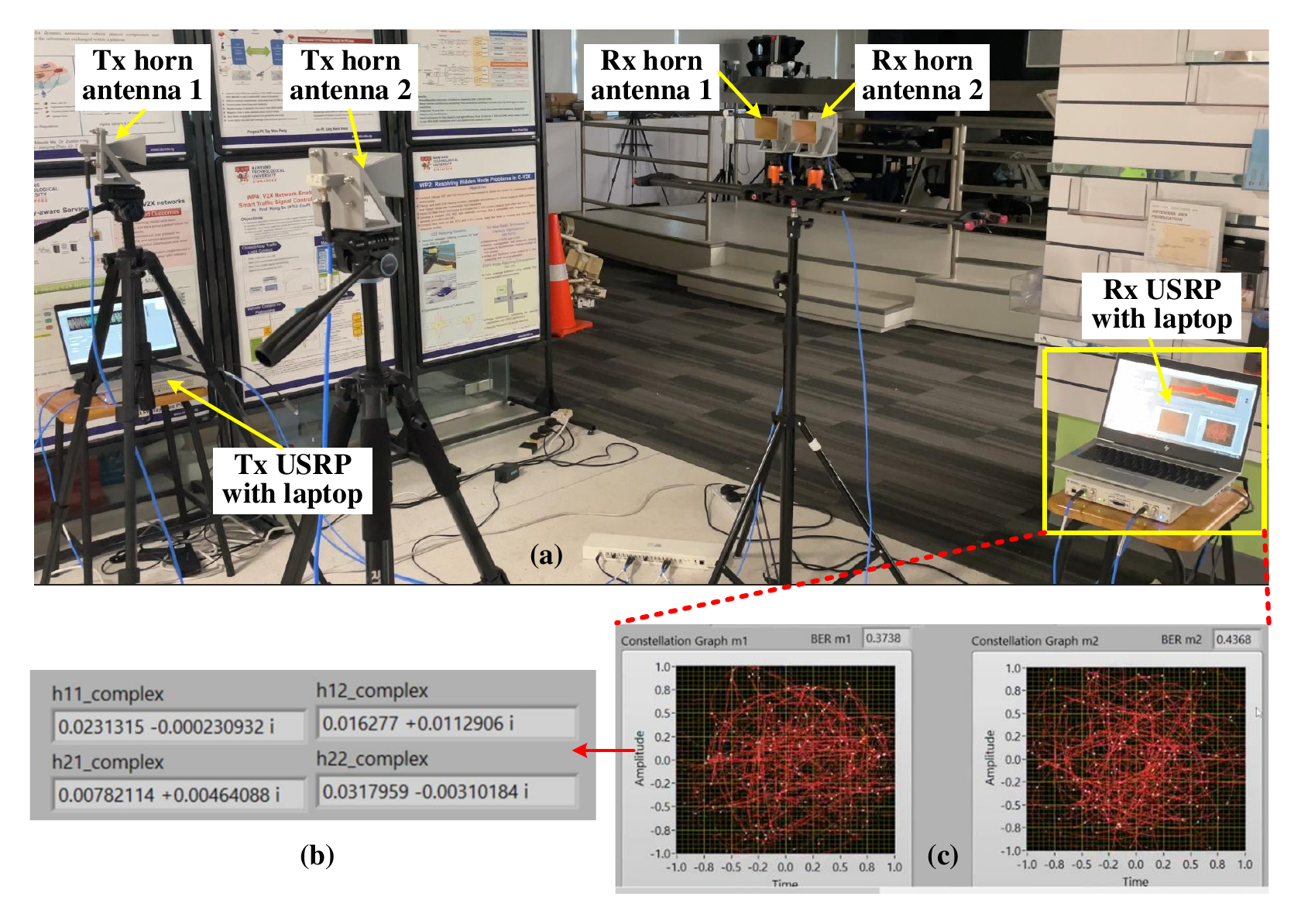}
\caption{Communications in high-correlation scenario with general horn antennas at Tx side. (a) Experiment setups. (b) Measured channel matrix. (c) Demodulated constellations.}
\label{horn-setup}
\end{figure}

Then, we switched the OAM transmitters to conventional planar wave transmitters, specifically 2 horn antennas, as depicted in Fig. \ref{horn-setup}. It is important to note that the antenna gain of the designed OAM antenna is approximately 16 dB, which is nearly identical to that of the horn antennas used in Fig. \ref{horn-setup}. During the experiments, the maximum gain directions of both types of antennas are aligned towards the receiving area. Hence, the receiving signals have the similar average power at the above two setups. The only difference comes from the channel correlation. Similarly, in this setting, the channel correlation coefficient of the $2 \times 2$ horn antennas transmission can be calculated as,
\begin{equation} \label{horn_correlation}
{\rho _{{\rm{horn}}}} = \frac{{\left| {{{h'}_{11}}{h'}_{21}^ *  + {{h'}_{12}}{h'}_{22}^ * } \right|}}{{\sqrt {(|{{h'}_{11}}{|^2} + |{{h'}_{12}}{|^2}) \times (|{{h'}_{21}}{|^2} + |{{h'}_{22}}{|^2})} }}{\rm{  =  }}0.7189.
\end{equation}
Comparing \eqref{OAM_correlation} and \eqref{horn_correlation}, it is evident that under the same conditions of Tx and Rx locations, power, and antenna gain, the channel correlation coefficient achieved with NTCS-OAM antennas is significantly better than that with traditional planar EM wave horn antennas, which benefits the multi-pair demodulation constellations. This improvement is due to the NTCS-OAM antennas' ability to modulate the wavefront phase additionally while ensuring a fixed main lobe pattern, thereby further reducing the correlation properties of LoS channels.

\begin{figure}[htbp]
\centering
\includegraphics[width=3.2in]{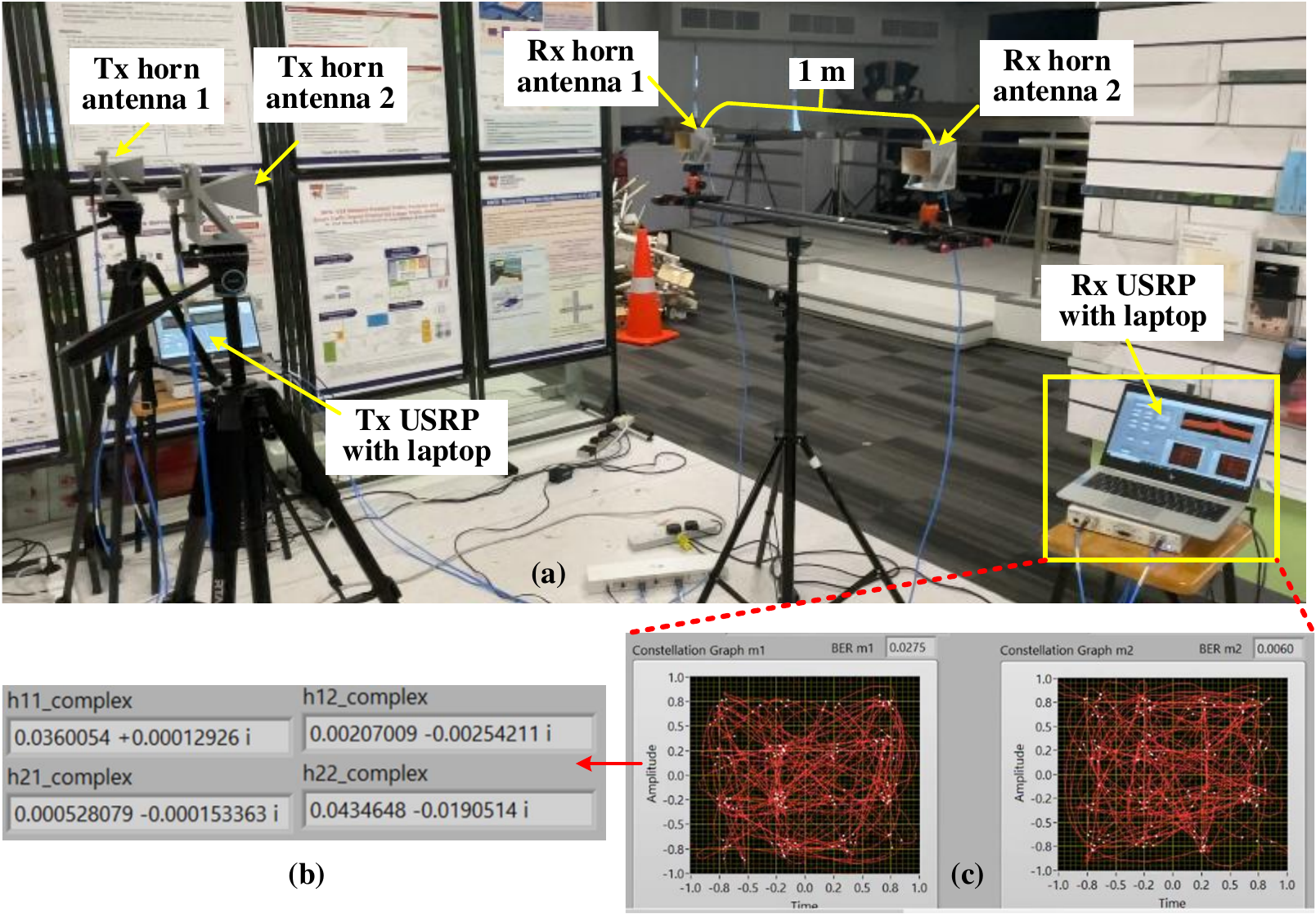}
\caption{Communications in low-correlation scenario with general horn antennas at Tx side. (a) Experiment setups. (b) Measured channel matrix. (c) Demodulated constellations.}
\label{Separate_Plane}
\end{figure}

Comparing \ref{OAM-setup}(c) and \ref{horn-setup}(c), it is obvious that due to the additional wavefront diversity, the demodulated constellation graphs of OAM multi-channel are much more clearer than the planar waves. These experimental results show that OAM beams can achieve significantly better communication performance than conventional planar waves in high-correlation channel scenarios. Furthermore, based on experiment settings in Fig. \ref{horn-setup}, we separate the Rx antennas with larger spacing, as shown in Fig. \ref{Separate_Plane}, which can be regarded as the low-correlation communication scenarios. It can be seen that in this case, the multiplexed signals can be successfully demodulated as shown in Fig. \ref{Separate_Plane}(c), and according to Fig. \ref{Separate_Plane}(b), the channel correlation coefficient is also reduced as,
\begin{equation} \label{horn_correlation_separate}
{{\bar \rho }_{{\rm{horn}}}} = \frac{{\left| {{{h''}_{11}}{h''}_{21}^ *  + {{h''}_{12}}{h''}_{22}^ * } \right|}}{{\sqrt {(|{{h''}_{11}}{|^2} + |{{h''}_{12}}{|^2}) \times (|{{h''}_{21}}{|^2} + |{{h''}_{22}}{|^2})} }}{\rm{ = }}0.0993.
\end{equation}
This further indicates that high channel correlation leads to significant interference between multiplexed data streams, making it difficult to separate and demodulate the multi-channel data. This interference adversely affects the overall capacity of the communication system.

\begin{figure}[htbp]
\centering
\includegraphics[width=3.2in]{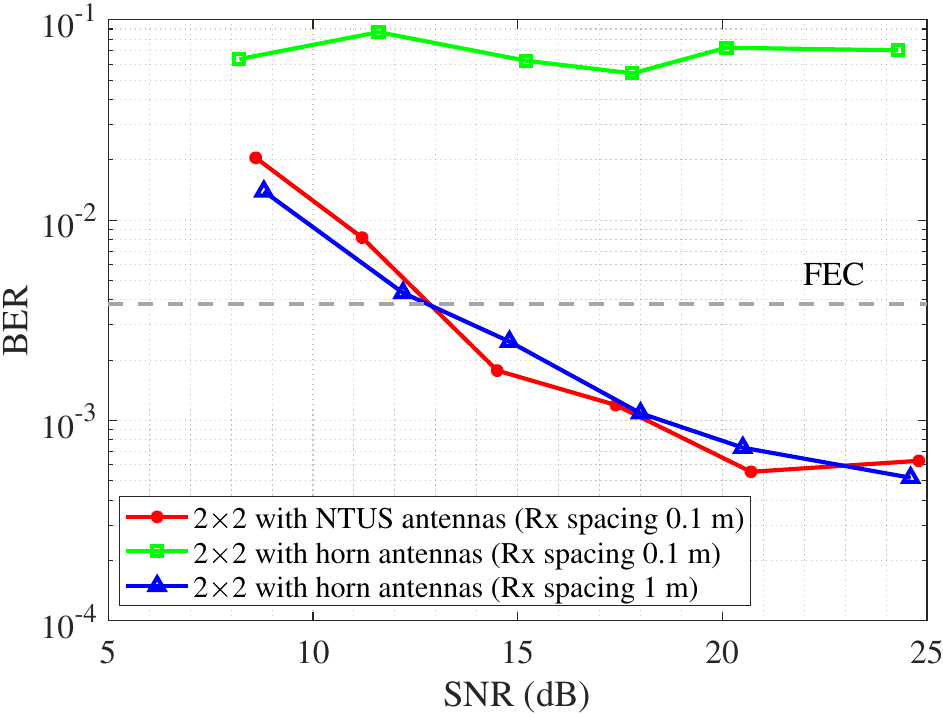}
\caption{The variation of BERs of demodulated signals versus SNR.}
\label{BER}
\end{figure}

Furthermore, to further verify the robustness of the experimental results under various noise conditions and eliminate the influence of random factors, we tracked how the Bit Error Rate (BER) varies with the SNR across different settings, as shown in Fig. \ref{BER}.
As expected, under the conditions of highly correlated LoS channels, the BER for ordinary horn antennas is the highest. Even at high SNR conditions, the BER remains significantly elevated, reaching $10^{-1}$, due to the interferences between multiple channels. However, when we replace the horn antennas with NTCS-OAM antennas, the BER quickly decreases due to the low correlation characteristics of OAM beams. At higher SNR, the BER can drop below the Forward Error Correction (FEC) threshold ($3.8 \times {10^{ - 3}}$) \cite{FEC}. This phenomenon is consistent with the improved BER performance obtained by increasing the spacing between the receiving antennas. This indicates that the NTCS-OAM antennas proposed in this paper can reduce the correlation between channels, which is equivalent to creating a greater physical spatial separation.

It is worth noting that all the above communication experiments were conducted without the use of pre-coding or channel equalization algorithms. The purpose of this approach is to eliminate the influence of other communication algorithms and focus solely on the impact of the channel in its raw transmission state. Indeed, the antenna design is a method of RF analog processing that does not impact the design of traditional communication algorithms. In other words, conventional baseband signal processing algorithms focus solely on the digital components beyond the RF interface, treating the antenna as a part of the generalized physical channel. The wavefront modulation method based on NTCS, as proposed in this paper, is an analog processing approach. This means it provides a new degree of freedom for actively changing the physical channel. In the future, we plan to explore more flexible wavefront control methods to pave the way for more intelligent communication systems.

The channel capacity improvements with the novel OAM beams are primarily due to the reduction in channel correlation coefficients. In LoS MIMO scenarios, high channel correlation limits multiplexing capabilities \cite{Ni}. The unique wavefront phase variations of the novel OAM beams introduce additional diversity, resulting in lower correlation between channels. This allows for more independent data streams to be transmitted simultaneously, enhancing spectral efficiency and overall channel capacity. By performing SVD of the MIMO channel, we obtained the singular value distribution, which reflects the system's multiplexing capability. Larger singular values receive higher power allocations, resulting in greater data throughput \cite{Chengnew,Lee}. Tables I and II show the channel condition numbers under different simulation settings, highlighting the changes in channel capacity. Our experiments demonstrated that the novel OAM beams provided a more orthogonal set of channels, leading to improved BER and signal-to-interference ratio.

It should be noted that the novel OAM beams proposed in this paper still retain the strict orthogonality characteristics between different modes. The orthogonal reception of these pencil-shaped OAM beams requires precise alignment conditions, which are theoretically feasible but demand meticulous geometric adjustments. For instance, achieving strict orthogonal transmission involves ensuring the wavefront phases of the OAM beams differ by an integer multiple of the wavelength at the receiving antennas. Besides the known orthogonality of OAM modes, we aimed to highlight that OAM beams, with their unique wavefront phase variations, can enhance the diversity of the physical channel. This is significant for reducing correlation between different channels, which is particularly beneficial for increasing channel capacity in LoS communication scenarios. Typically, LoS channels exhibit strong correlations due to the lack of multipath-induced randomness. However, if we consider the transmitting antenna as part of the physical channel, introducing additional EM variations via OAM, this can bring extra diversity to the otherwise monotonous LoS channel. Specifically, by introducing wavefront phase variations at the transmitting antennas while maintaining similar transmission power and radiation patterns, OAM can effectively reduce channel correlation in the EM space. This also presents a potential degree of freedom for future beamforming and precoding designs.

\section{Conclusion and Discussion}
This paper proposes a novel LoS-MIMO communication architecture based on non-traditional OAM beams, challenging the conventional understanding of inherently cone-shaped OAM beams. By innovating the design of the OAM transmitter, we have generated OAM beams with directional emission devoid of central energy voids, whose radiation patterns now closely resemble those of traditional planar wave horn antennas. Excitingly, within the main lobe of the antenna's radiation pattern, we cleverly preserved the phase variation characteristics of the OAM beams, linking different modes to the linear wavefront variation gradients within the main lobe. To validate the effectiveness of this design, we bridged theory and practice by constructing an actual LoS-MIMO communication experimental platform, measuring the channel correlation coefficients, communication stability, and BER characteristics under various scenario settings. This constitutes a robust piece of research work that offers both academic and engineering guidance for the construction of practical communication infrastructures. To address multiplexing, our design includes mounting plates and screw holes for future assembly of different mode waveguides, allowing combined multi-mode transmission. Although the current design is bulky, other compact implementations like PCB antennas and U-shaped waveguide resonators, which are also good options. In future research, we will continue to explore OAM multiplexing methods while reducing the size and cost of the OAM transmitters.

As usual, the novel pencil-shaped OAM beams proposed in this paper can achieve orthogonality between modes, making them suitable for existing point-to-point communication frameworks and potentially supporting longer-distance communication due to the absence of energy divergence. In this application scenario, precise alignment is necessary to distinguish between modes. Additionally, it should be noted that unlike traditional OAM beams, these new beams have radiation patterns similar to planar wave antennas, allowing integration into 5G and future 6G mobile and non-LoS communication scenarios, where the precise alignment requirement may not be necessary. The unique, adjustable wavefront phase characteristics introduce additional variations in the physical channel, enhancing channel capacity and providing a new degree of freedom for communication algorithm design. We believe that with advancements in reconfigurable antenna technology, these OAM beams will offer exciting possibilities as we transition from 5G to 6G.

\end{document}